\documentclass[12pt]{article}
\usepackage[dvips]{graphicx}
\usepackage{epsfig}

\begin{document}

\title{\Large \bf High precision study of $K^\pm\to3\pi^\pm$ decays by NA48/2}
\author{\large Evgueni Goudzovski\footnote{On behalf of the NA48/2
collaboration: Cambridge-CERN-Chicago-Dubna-Edinburgh-Ferrara-Firenze-%
Mainz-Northwestern-Perugia-Pisa-Saclay-Siegen-Torino-Wien.}\bigskip \\
{\it Joint Institute for Nuclear Research, Dubna, 141980 Russia}}

\maketitle

{\large

\begin{abstract}
\medskip
Preliminary results of study of $K^\pm\to3\pi^\pm$ decays
by the NA48/2 experiment at CERN SPS are presented.
They include a precise measurement of the
direct CP violating charge asymmetry of Dalitz plot linear slope
parameters $A_g=(g^+-g^-)/(g^++g^-)$, and
a measurement of the Dalitz plot slope parameters
$(g,h,k)$ themselves. Due to the design of the experiment,
and a large data set collected, unprecedented precisions were
achieved.
\end{abstract}

\vspace{-9mm}
\section*{Introduction}
\vspace{-2mm}

More than 40 years after its discovery~\cite{ch64}, the phenomenon
of CP violation (CPV) plays a central role in
particle physics investigations. For a long time it seemed
to be confined to a peculiar sector of particle physics. However,
two breakthroughs took place recently. In the late 1990s, following
an earlier indication by NA31~\cite{ba93}, the NA48 and KTeV
experiments firmly established the existence of direct CPV~\cite{fa99,al99}
by measuring a non-zero
$\varepsilon'/\varepsilon$ parameter in $K^0\to2\pi$ decays.
In the early 2000s, a series of indirect~\cite{au01} and
direct~\cite{ab04} CPV effects in $B$ meson
decays was discovered.
In kaon physics, the charge asymmetry between
$K^+$ and $K^-$ decays into $3\pi$ discussed here
is among the most promising observables, along with the
parameter $\varepsilon'/\varepsilon$, and rates
of GIM-suppressed FCNC decays $K\to\pi\nu\bar\nu$.

The $K^\pm\to3\pi^\pm$ matrix element squared is conventionally
parameterized by a polynomial expansion~\cite{pdg}
\begin{equation}
|M(u,v)|^2\sim C(u,v)\cdot(1+gu+hu^2+kv^2), \label{slopes}
\end{equation}
where $g$, $h$, $k$ are the linear and quadratic Dalitz
plot slope parameters ($|h|,|k|\ll |g|$), $C(u,v)$ is the
Coulomb factor\footnote{$C(u,v)=\prod_{i,j=1,2,3;i\ne j}
\{n_{ij}/(e^{n_{ij}}-1)\}$,~~$n_{ij}=2\pi\alpha e_ie_j/\beta_{ij}$,~~where
$e_i=\pm1$ -- pion charges, $\beta_{ij}$ -- relative velocities
of pion pairs.}, and the two
kinematic variables $u$ and $v$ are defined as
\begin{equation}
u=\frac{s_3-s_0}{m_\pi^2},~v=\frac{s_2-s_1}{m_\pi^2},~
s_i=(P_K-P_i)^2,~i=1,2,3;~s_0=\frac{s_1+s_2+s_3}{3}.
\end{equation}
Here $m_\pi$ is the charged pion mass, $P_K$ and $P_i$ are the kaon
and pion four-momenta, the indices $i=1,2$ correspond to the two
identical (``even'') pions and the index $i=3$ to the other
(the ``odd'') pion. A term proportional to $v$ is forbidden
in (\ref{slopes}) by Bose symmetry. A difference of slope
parameters $g^+$ and $g^-$ describing positive and negative kaon
decays, respectively, is a manifestation of direct CPV
usually defined by the linear slope asymmetry
\begin{equation}
A_g = (g^+ - g^-)/(g^+ + g^-) \approx \Delta g/(2g),
\end{equation}
where $\Delta g$ is the slope difference and $g$ is the average
slope. The asymmetry of decay rates $A_\Gamma$ is expected to be
strongly suppressed with respect to $A_g$~\cite{is92}.

Several experiments had searched for the asymmetry $A_g$ in both
$\pi^\pm\pi^+\pi^-$~\cite{fo70} and $\pi^\pm\pi^0\pi^0$~\cite{sm75}
$K^\pm$ decay modes before the NA48/2. Resulting upper limits are at the
level of $10^{-3}$, with large contributions of systematic uncertainties.
This precision is unsatisfactory, since the SM predictions for $A_g$
vary from a few $10^{-6}$ to a few
$10^{-5}$~\cite{ma95}, while existing calculations involving
processes beyond the SM~\cite{sh98} predict
enhancements up to a few $10^{-4}$.
The primary goal of the NA48/2 experiment is $A_g$ measurement
in both decay modes with a new level of precision of $10^{-4}$
using a technique of simultaneous $K^+/K^-$ beams,
thus significantly reducing the gap between
experiment and theory. NA48/2 $K^\pm\to3\pi^\pm$ results
obtained with partial data sample were published~\cite{ag}.

Precise study of $K^\pm\to3\pi^\pm$ Dalitz plot distribution is of interest
as such, since it has been recently demonstrated experimentally
by study of $K^\pm\to\pi^\pm\pi^0\pi^0$ decay distribution by NA48/2~\cite{cusp}
and subsequently understood theoretically~\cite{ca05} that, due to
effects of final state pion rescattering, the density of
$K_{3\pi}$ Dalitz plot can be used to measure the $\pi\pi$ scattering lengths.

\vspace{-5mm}
\section{Beams and detectors}
\vspace{-2mm}

High precision measurement of $A_g$
requires a dedicated experimental approach alongside with collection
of large data samples. A novel beam line providing two simultaneous
charged beams of opposite signs overlapping in space was designed
and built in ECN3 high intensity hall at the CERN SPS.
Allowing decays of $K^+$ and $K^-$ to be recorded at the same time,
it serves as a key element of the experiment, leading to
cancellations of main systematic uncertainties. Regular alternation
of magnetic fields in all the beam line elements was adopted to
symmetrize the acceptance for the two beams. The layout of the beams
and detectors is shown schematically in Fig.~\ref{fig:beams}.

The setup is described in a right-handed coordinate
system with the $z$ axis directed downstream along the beam, and the
$y$ axis directed vertically up.

The beams are produced by 400 GeV protons impinging on a beryllium
target at zero angle.
Charged particles with momentum $(60\pm3)$ GeV/$c$ are selected
in a charge-symmetric way by an achromatic
system of four dipole magnets with null total deflection, which
splits the two beams in the vertical plane and recombines them
on a common axis. Then the beams pass through a
series of 4 quadrupoles designed to produce
charge-symmetric focusing of the beams towards the
detector. Finally they are again split and recombined in a second
achromat housing a kaon beam spectrometer.

\begin{figure}[tb]
\vspace{-3mm}
\begin{center}
{\resizebox*{0.82\textwidth}{!}{\includegraphics{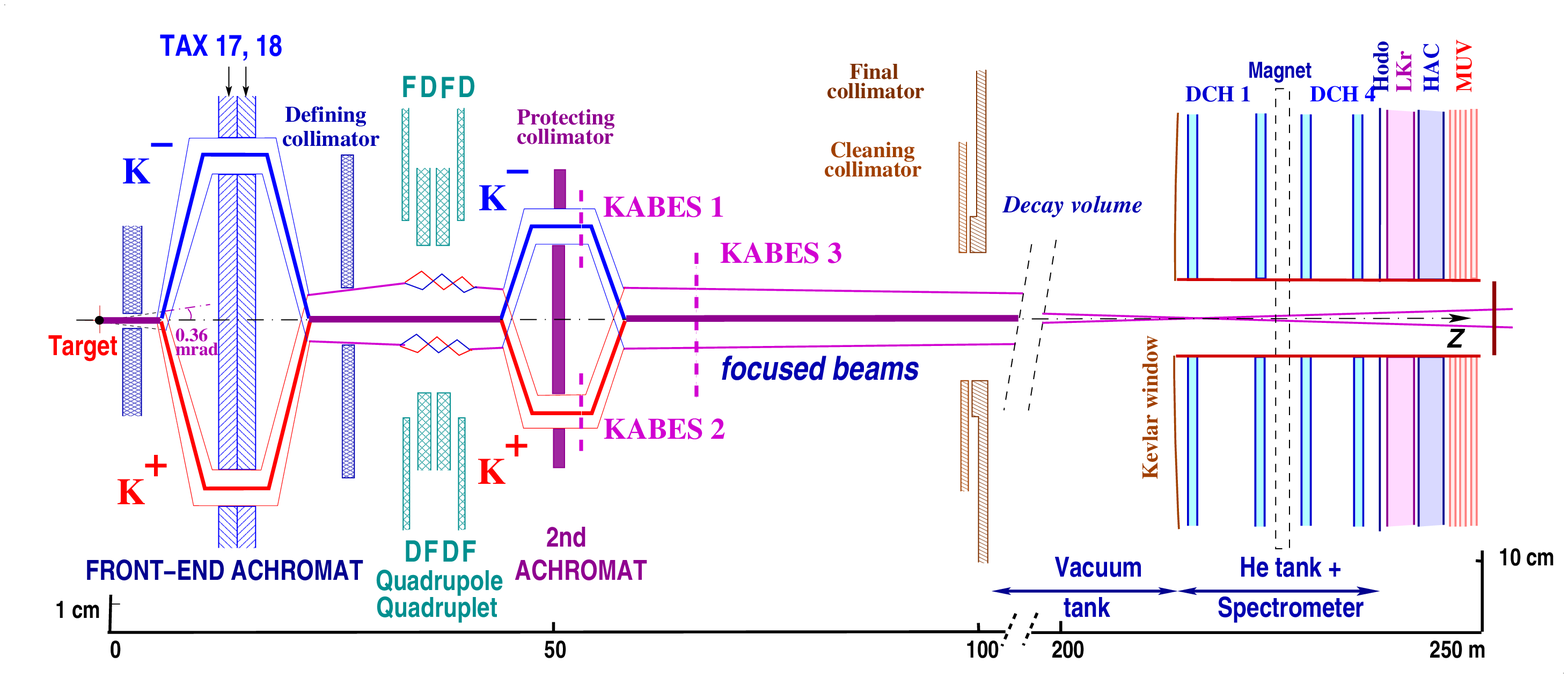}}}
\end{center}
\vspace{-11mm} \caption{A view of NA48/2 beam
line (TAX17,18: motorized beam dump selecting
momentum of $K^+/K^-$ beams; DFDF: focusing quadrupoles,
KABES1--3: beam spectrometer) and
detector (DCH1--4: drift chambers, Hodo: hodoscope, LKr: EM
calorimeter, HAC: hadron calorimeter, MUV: muon veto). Vertical
scale differs in two parts of the figure.}
\vspace{-5mm}
\label{fig:beams}
\end{figure}

Further downstream both beams follow the same path.
After passing the cleaning and the final collimators they enter the
decay volume, housed in a 114m long cylindrical vacuum tank. With
$7\times 10^{11}$ protons per burst of 4.8s duration
incident on the target, the positive (negative) beam flux at the
entrance of the decay volume is $3.8\times 10^7$ ($2.6\times 10^7$)
particles per pulse, of which $5.7\%$ ($4.9\%$) are $K^+$ ($K^-$).
The $K^+/K^-$ flux ratio is about $1.8$. The fraction of beam kaons
decaying in the decay volume is about $22\%$.

The decay volume is followed by a magnetic spectrometer used for the
reconstruction of $K^\pm\to3\pi^\pm$ decays. It is
housed in a tank filled with helium at atmospheric pressure,
separated from the vacuum tank by a thin ($0.31\%X_0$) {\it Kevlar}-composite
window. A thin-walled aluminium
tube of 16 cm diameter traversing the centres
of all detectors allows the undecayed beam
particles to continue
their path in vacuum. The spectrometer consists of 4 drift
chambers (DCH): two located upstream and two downstream of the
dipole magnet providing a horizontal transverse momentum kick
of 120 MeV/$c$ to the charged particles. The DCHs have a shape
of a regular octagon with transverse size of about 2.8 m.
Each DCH is composed of 8
planes of sense wires arranged in 4 couples of staggered planes
oriented horizontally, vertically, and along each
of the two orthogonal $45^\circ$ directions. Momentum resolution
of the spectrometer is $\sigma_p/p = 1.02\% \oplus
0.044\%p$ ($p$ in GeV/$c$), corresponding to $3\pi$
invariant mass resolution of
$1.7$~MeV/$c^2$.

The spectrometer is followed by a scintillator hodoscope (HOD)
consisting of a plane of horizontal and a plane of vertical strips,
64 strips arranged in 4 quadrants in each plane.
A liquid krypton EM calorimeter
(LKr), a hadronic calorimeter (HAC), and a muon detector (MUV)
follow downstream.

The $K^\pm\!\to\!3\pi^\pm$ decays are triggered with a two-level system.
At the first level (L1), the rate of $\sim500$ kHz is reduced to
$\sim100$ kHz by requiring coincidences of hits in the two HOD planes
in at least two quadrants. The second
level (L2) is based on a hardware system computing coordinates of
DCH hits using drift times, and a farm of asynchronous
microprocessors performing fast reconstruction of tracks and running
the decision taking algorithm. The L2 algorithm requires at least
two tracks to originate in the decay volume with the closest
distance of approach of less than 5 cm. L1 triggers not satisfying
this condition are examined further and accepted if there is a
reconstructed track that is not compatible with a
$\pi^\pm\pi^0$ decay of a $K^\pm$ having 60 GeV/$c$ momentum
directed along the $z$ axis. The resulting trigger rate is $\sim10$
kHz.

NA48/2 collected data during two runs in 2003 and 2004, with
$\sim$50 days of efficient data taking each. About $18\times 10^9$
triggers were totally recorded.
\vspace{-5mm}

\section{Asymmetry measurement method}
\vspace{-2mm}

The measurement method is based on comparing the reconstructed $u$
spectra of $K^+$ and $K^-$ decays $N^+(u)$ and $N^-(u)$. Given the
slope parameters values ~\cite{pdg}
and the precision of the measurement, the ratio of $u$
spectra of $R(u)=N^+(u)/N^-(u)$ is in good approximation
proportional to $(1+\Delta g\cdot u)$, so $\Delta g$ can be
extracted from a linear fit to $R(u)$, and $A_g=\Delta
g/2g$ can be evaluated.

Charge symmetrization of experimental conditions is to a large
extent achieved by using simultaneous collinear $K^+/K^-$
beams of similar momentum spectra. However, the presence of
magnetic fields in both the beam line (achromats,
quadrupoles) and the spectrometer, combined with some
asymmetries in detector performance, introduces residual charge
asymmetries. To equalize local effects on the
acceptance, polarities of all the magnets in the beam line were
reversed during the data taking on an approximately weekly basis
(corresponding to the periodicity of SPS technical stops), while the
polarity of the spectrometer magnet was alternated on a more
frequent basis.

Data collected over a period with all the four possible setup
configurations (i.e. combinations of beam line and spectrometer
magnet polarities) spanning about two weeks of data taking
represent a ``supersample'', which is treated as an independent
self-consistent set of data for asymmetry measurement. Nine
supersamples numbered 0 to 8 were collected in two years of running.

Each supersample contains eight distinct data samples
corresponding to various combinations of setup configurations
and kaon sign. In order to minimize the effects of beam and detector
asymmetries, the following
``quadruple ratio'' involving the eight $u$ spectra,
composed as a product of four $R(u)=N^+(u)/N^-(u)$ ratios with
opposite kaon sign, and deliberately chosen setup configurations in
numerator and denominator, is considered:
\begin{equation}
R_4(u) = R_{US}(u)\cdot R_{UJ}(u)\cdot R_{DS}(u)\cdot R_{DJ}(u).
\label{quad}
\end{equation}
Here the indices $U$ ($D$) denote
beam line polarities corresponding to $K^+$ passing along the
upper (lower) path in the achromats, and the indices
$S$ ($J$) represent spectrometer magnet polarities (opposite for
$K^+$ and $K^-$) corresponding to the ``even''
 pions deflection to negative (positive) $x$, i.e.
towards the Sal\`eve (Jura) mountains. Fitting
the ratio~(\ref{quad}) with $f(u)=n\cdot(1+\Delta
g\cdot u)^4$ results in two parameters: the normalization $n$ and
the slope difference $\Delta g$.

The quadruple ratio technique completes the procedure of
polarity reversals, and allows a three-fold cancellation of
systematic biases: 1) beam line biases cancel between $K^+/K^-$ samples
with the beams following the same path; 2) local detector
biases cancel between $K^+/K^-$ samples with decay products
illuminating the same parts of the detector;
3) due to simultaneous $K^+/K^-$ beams,
global time-variable biases cancel
between $K^+/K^-$ samples.

The method is independent of the $K^+/K^-$ flux ratio and the
relative sizes of the samples collected with different setup
configurations. The result remains sensitive only
to time variations of asymmetries in the experimental conditions
with characteristic times smaller than corresponding field
alternation period, and in principle should be free of systematic
biases.

With the above method, no Monte Carlo (MC) corrections
to the acceptance are required. Nevertheless, a detailed GEANT-based
MC simulation was developed as a tool for systematic studies,
including full geometry and material description,
simulation of time variations of local DCH inefficiencies,
beam geometry and DCH alignment. A large MC
production was made, providing a sample of a size comparable
to that of the data ($\sim 10^{10}$ events).

\vspace{-5mm}
\section{Asymmetry measurement}
\vspace{-2mm}

Tracks are reconstructed from hits in DCHs using the measured
magnetic field map of the spectrometer analyzing magnet rescaled
according to the recorded current. Three-track vertices compatible
with a $K^\pm\to3\pi^\pm$ decay topology are reconstructed by extrapolation
of track segments from the spectrometer upstream
into the decay volume, taking into account the stray magnetic fields
in the decay volume, and multiple
scattering at the Kevlar window.

Event selection includes
requirements on vertex charge, quality, and position (within
the decay volume, laterally within the beam), limits on the
reconstructed $3\pi$ momentum: $54~{\rm GeV}/c<P_K<66~{\rm GeV}/c$
and invariant mass: $|M_{3\pi}-M_K|<9$~MeV/$c^2$ (the latter
corresponding to five times the resolution). The
selection leaves a practically background free sample,
as $K_{3\pi}$ is the dominant three-track $K^\pm$ decay mode.

{\bf\underline{Fine alignment of the DCHs.}}
Transverse positions of DCHs and individual wires were
realigned every 2--4 weeks of data taking with a precision of
30$\mu$m using data collected in special runs.
However, time variations of DCH
alignment on a shorter time scale can potentially bias the
asymmetry, since an uncorrected shift of a DCH along the $x$ axis
leads to charge-antisymmetric mismeasurement of the momenta.  An
unambiguous measure of the residual misalignment is the difference
between the average reconstructed $3\pi$ invariant masses for $K^+$
and $K^-$ decays ($\Delta\overline M$).
Monitoring of $\Delta \overline M$ revealed significant (up
to 200 $\mu$m) movements of the DCHs between individual alignment
runs. Introduction of time-dependent corrections to the measured
momenta based on the observed $\Delta\overline M$ reduces the effect
on the slope difference by more than an order of magnitude to a
negligible level of $\delta(\Delta g)<0.1\times 10^{-4}$.

{\bf\underline{Correction for beam geometry instabilities.}}
The main feature determining the geometric
acceptance is the beam pipe traversing the centres of DCHs.
Beam optics control transverse beam
positions only to $\pm1$~mm, leaving a sizable random charge-asymmetric
bias to the acceptance. To compensate for this effect,
inner radius cuts $R>11.5$~cm are introduced for the distances
of pion impact points at the first and the last DCHs from the actual
average measured beam positions, at the cost of $12\%$ of the
statistics.
The minimum distance of 11.5 cm is chosen to ensure that the region
of the beam tube and the adjacent central insensitive areas of the
DCHs are securely excluded by the cut. The average beam positions
are continuously monitored separately for $K^+$ and $K^-$.
In addition to
the time variation of the average positions, also the dependencies
of the beam position on kaon momentum
are monitored and taken into account. A precision of
$100~\mu$m in the determination of beam position is sufficient to
reduce systematic effects to a negligible level.
Residual charge-asymmetric effects originate from permanent
magnetic fields in the decay volume coupling to
time-dependent DCH inefficiencies and beam migrations. The
corresponding fake asymmetries do not exceed
$\delta(\Delta g)=0.2\times10^{-4}$.

{\bf\underline{Trigger efficiency correction.}}
Only charge-asymmetric inefficiencies dependent on
$u$ can bias the measurement. Inefficiencies of trigger
components are measured as functions of $u$ using control data
samples from low bias triggers collected along with the main
triggers, which allows to account for their time variations,
and propagate their statistical errors into the result.

L1 trigger inefficiency, due to hodoscope inefficiency,
was measured to be $0.9\times 10^{-3}$ and stable in
time. Due to its time stability, no
correction is applied, and an uncertainty of $\delta(\Delta
g)=0.3\times 10^{-4}$, limited by the size of the control
sample, is attributed.
For the L2 trigger, corrections to $u$ spectra are introduced for
the rate-independent part of the inefficiency, which is
time-dependent due to instabilities of the local DCH inefficiencies
affecting the trigger more than the reconstruction.
The integral inefficiency for the selected sample is
normally close to $0.6\times 10^{-3}$. The correction
 amounts to $\delta(\Delta g)\!=\!(-0.1\pm0.3)\!\times\!
10^{-4}$; its the error is statistical due to limited size of
the control sample. The symmetry of the rate-dependent part of the
inefficiency of $\sim 0.2\%$ was checked separately with MC
simulation of pile-up effects.

{\bf\underline{Asymmetry fits and cross-checks.}}
After applying the above corrections, $\Delta g$
is extracted by fitting the quadruple ratio of the $u$ spectra
(\ref{quad}) for each supersample. Statistics
selected in each supersample, the ``raw'' values of $\Delta
g$, and the final
values of $\Delta g$ with the L2 trigger corrections applied are
presented in Table~\ref{tab:stats}. The independent results obtained
in the nine supersamples are shown in Figure \ref{fig:stabplot}(a):
they are compatible with $\chi^2/{\rm ndf}
= 10.0/8$.

\begin{table}[tb]
\begin{center}
\begin{tabular}{r|r|r|r|r}
\hline Supersample & $K^+\!\to\pi^+\pi^+\pi^-\!\!$ &
$K^-\!\to\pi^-\pi^-\pi^+\!\!$
&$\Delta g\times 10^4$ & $\Delta g\times 10^4$\\
&decays in $10^6\!\!$&decays in $10^6\!\!$&raw~~~~~&corrected\\
\hline
0 & 448.0 & 249.7 & $ 0.5\pm1.4$ & $-0.8\pm1.8$\\
1 & 270.8 & 150.7 & $-0.4\pm1.8$ & $-0.5\pm1.8$\\
2 & 265.5 & 147.8 & $-1.5\pm2.0$ & $-1.4\pm2.0$\\
3 &  86.1 &  48.0 & $ 0.4\pm3.2$ & $ 1.0\pm3.3$\\
4 & 232.5 & 129.6 & $-2.8\pm1.9$ & $-2.0\pm2.2$\\
5 & 142.4 &  79.4 & $ 4.7\pm2.5$ & $ 4.4\pm2.6$\\
6 & 193.8 & 108.0 & $ 5.1\pm2.1$ & $ 5.0\pm2.2$\\
7 & 195.9 & 109.1 & $ 1.7\pm2.1$ & $ 1.5\pm2.1$\\
8 & 163.9 &  91.4 & $ 1.3\pm2.3$ & $ 0.4\pm2.3$\\
\hline
Total & 1998.9 & 1113.7 & $0.7\pm0.7$ & $0.6\pm0.7$\\
\hline
\end{tabular}
\end{center}
\vspace{-7mm} \caption{Selected statistics
and measured $\Delta g$ in each supersample.}
\vspace{-4mm}
\label{tab:stats}
\end{table}

\begin{figure}[tb]
\vspace{-1mm}
\begin{center}
\resizebox{0.74\textwidth}{!}{\includegraphics{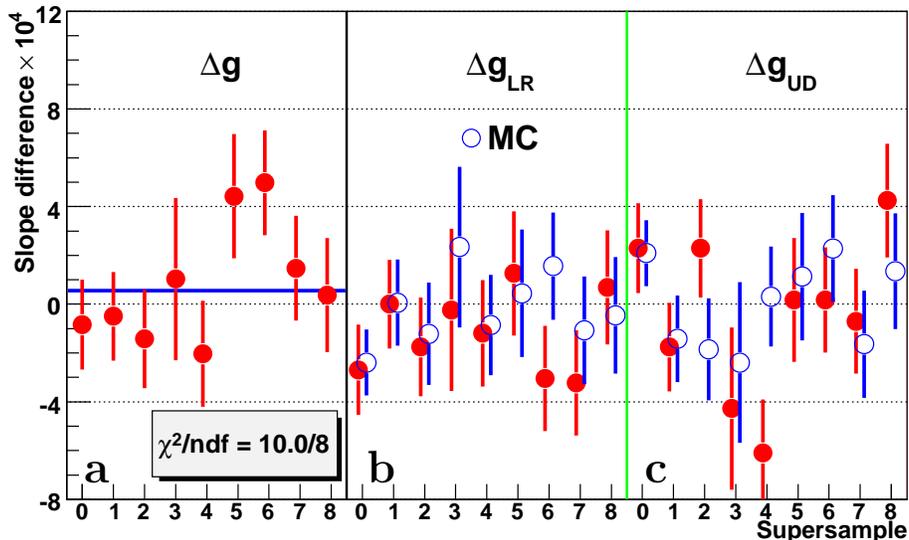}}
\put(-317,28){\Large\bf a} \put(-210,28){\Large\bf b}
\put(-105,28){\Large\bf c}
\end{center}
\vspace{-11mm} \caption{(a) $\Delta g$ measurement in the four
supersamples; control
  quantities (b) $\Delta g_{LR}$ and (c) $\Delta g_{UD}$ corresponding
  to detector and beam line asymmetries which cancel in quadruple
  ratio, and their comparison to Monte Carlo.}
\vspace{-6mm}
\label{fig:stabplot}
\end{figure}

To measure sizes of the systematic biases
cancelling due to the quadruple ratio technique, two other quadruple
ratios of the eight $u$ spectra were formed. These are the products
of four ratios of $u$ spectra of same sign kaons recorded with
different setup configurations, therefore any physical asymmetry
cancels in these ratios, while the setup asymmetries do not.
The corresponding fake slope differences $\Delta g_{LR}$ and $\Delta
g_{UD}$ in the nine supersamples are presented in Figure
\ref{fig:stabplot}(b) and (c) for both data and MC. The size of
these control quantities demonstrates that the cancellation of the
first-order systematic biases in (\ref{quad}) is at the level of a
few $10^{-4}$; therefore second order effects are negligible.

{\bf\underline{Limits for residual systematic effects.}}
Pion momentum measurement is based on the knowledge
of spectrometer magnet magnetic field. The precision of
magnetic field reversal is $10^{-3}$.
Imperfect inversion effect is charge symmetric due to the
simultaneous beams. An upper limit for the corresponding
systematic uncertainty is $\delta(\Delta g) = 0.1\times 10^{-4}$.
In a considerable fraction of the selected events ($\sim 5\%$)
a pion undergoes a $\pi\to\mu\nu$ decay in the decay
volume. These
events dominate the tails of the reconstructed $3\pi$ mass distribution.
By varying the accepted $3\pi$ invariant mass interval in a
wide range (5--25 MeV/$c^2$), a conservative systematic uncertainty
of $\delta(\Delta g)=0.4\times 10^{-4}$ was attributed to effects
due to pion decays.
Effects of accidental activity and pile-up were extensively studied
with MC simulation, and found to be
charge-symmetric to a level of $\delta(\Delta g)=0.2\times 10^{-4}$,
limited by MC statistics.
Biases due to resolution effects were studied by varying
methods of $u$ variable calculation and binning.
The result is stable within $\delta(\Delta g)=0.3\times 10^{-4}$.
Charge-asymmetric material effects were found to be negligible.
A summary of the systematic uncertainties is
presented in Table~\ref{tab:syst}.

\begin{table}[tb]
\begin{center}
\begin{tabular}{l|c}
\hline
Systematic effect & Correction, uncertainty $\delta(\Delta g)\times 10^4$\\
\hline
Spectrometer alignment        & $\pm0.1$\\
Acceptance and beam geometry  & $\pm0.2$\\
Momentum scale                & $\pm0.1$\\
Pion decay                    & $\pm0.4$\\
Pile-up                       & $\pm0.2$\\
Resolution and fitting        & $\pm0.3$\\
\hline
Total systematic uncertainty  & $\pm0.6$\\
\hline
Level 1 trigger               & $\pm0.3$\\
Level 2 trigger               & $0.1\pm0.3$\\
\hline
\end{tabular}
\end{center}
\vspace{-6mm} \caption{Systematic uncertainties and
  correction for level 2 trigger inefficiency.}
  \vspace{-4mm}
\label{tab:syst}
\end{table}

\vspace{-5mm}
\section{Measurement of slope parameters}
\vspace{-2mm}

The experiment was designed for asymmetry measurements,
which are MC-independent analyses relying on cancellations of
systematic effects. Nevertheless, huge statistics and well tuned MC
allow also precise study of $K^\pm\to3\pi^\pm$ kinematic distribution.
The ultimate goal is a full description with factorization of pion rescattering
effects~\cite{ca05} and radiative corrections. However, the first
stage of the analysis is interpretation in the framework of the
polynomial parameterization (\ref{slopes}), and verification of
(\ref{slopes}) at the new level of precision.

The measurement method is based on fitting the reconstructed data
distribution in $(u,v)$ with a sum of 4 MC components corresponding
to the 4 terms in the polynom in (\ref{slopes}). The relative weights
of the 4 components corresponding to the best data/MC
shape agreement define the values of $(g,h,k)$. Supersamples
1--3 only are used for the preliminary analysis. Reconstruction,
selection, and correction procedures are identical to those described
in section 3. Uncertainties arise from spectrometer alignment and momentum
scale, precision of description of DCH resolution and kaon momentum spectrum,
trigger inefficiencies, and limited sizes of data and MC samples.

Validity of the parameterization (\ref{slopes}) was demonstrated with high
precision, no evidence for higher order terms was found.

\vspace{-5mm}
\section*{Preliminary results and conclusions}
\vspace{-2mm}

The difference in $K^\pm\to3\pi^\pm$ Dalitz plot slope parameter
is found to be
$$
\Delta g = g^+-g^- = (0.6 \pm 0.7_{stat.} \pm 0.4_{trig.} \pm
0.6_{syst.})\times 10^{-4},
$$
leading to a CPV charge asymmetry using
$g=-0.2154\pm0.0035$~\cite{pdg}:
$$
A_g = (-1.3 \pm 1.5_{stat.} \pm 0.9_{trig.} \pm 1.4_{syst.})\times
10^{-4} = (-1.3\pm2.3)\times 10^{-4},
$$
which does not contradict the SM,
and due to high precision can be used to constrain
SM extensions predicting enhancements of the
charge asymmetry.

The following values of $K^\pm\to3\pi^\pm$
Dalitz plot slope parameters were measured
at the first stage of Dalitz plot shape analysis:
$$
g\!=\!(-21.131\pm0.015)\%,~~
h\!=\!(1.829\pm0.040)\%,~~
k\!=\!(-0.467\pm0.012)\%.
$$
They are in agreement with the world average~\cite{pdg}
obtained by experiments made back in 1970s,
but are an order of magnitude better in precision.


\end{document}